\documentclass[usenatbib]{emulateapj}
\usepackage{graphicx}
\usepackage[flushleft]{threeparttable}
\usepackage[usenames,dvipsnames,svgnames,table]{xcolor}
\usepackage{amsmath,amssymb}
\usepackage{amsmath}

\usepackage{natbib}
\begin{document}

\def\etal{et al.\ \rm}
\def\ba{\begin{eqnarray}}
\def\ea{\end{eqnarray}}
\def\etal{et al.\ \rm}
\def\Fdw{F_{\rm dw}}
\def\Tex{T_{\rm ex}}
\def\Fdis{F_{\rm dw,dis}}
\def\Fnu{F_\nu}
\def\FJ{F_J}


\title{Radial Transport and Meridional Circulation in Accretion Disks}

\author{Alexander A. Philippov\altaffilmark{1,4} and Roman R. Rafikov\altaffilmark{2,3}}

\altaffiltext{1}{Department of Astrophysical Sciences, 
Princeton University, Ivy Lane, Princeton, NJ 08540}
\altaffiltext{2}{Institute for Advanced Study, Einstein Drive, Princeton, NJ 08540}
\altaffiltext{3}{Centre for Mathematical Sciences, Department of Applied Mathematics and Theoretical Physics, University of Cambridge, Wilberforce Road, Cambridge CB3 0WA, UK}
\altaffiltext{4}{sashaph@princeton.edu}


\begin{abstract}
Radial transport of particles, elements and fluid driven by internal stresses in three-dimensional (3D) astrophysical accretion disks is an important phenomenon, potentially relevant for the outward dust transport in protoplanetary disks, origin of the refractory particles in comets, isotopic equilibration in the Earth-Moon system, etc. To gain better insight into these processes, we explore the dependence of meridional circulation in 3D disks with shear viscosity on their thermal stratification, and demonstrate strong effect of the latter on the radial flow. Previous locally isothermal studies have normally found a pattern of the radial outflow near the midplane, switching to inflow higher up. Here we show, both analytically and numerically, that a flow, which is inward at all altitudes, is possible in disks with entropy and temperature steeply increasing with height. Such thermodynamic conditions may be typical in the optically thin, viscously heated accretion disks. Disks in which these conditions do not hold should feature radial outflow near the midplane, as long as their internal stress is provided by the shear viscosity. Our results can also be used for designing hydrodynamical disk simulations with a prescribed pattern of the meridional circulation.
\end{abstract}

\keywords{accretion, accretion disks --- protoplanetary disks --- hydrodynamics }



\section{Introduction}  
\label{sect:intro}


Evolution of astrophysical accretion disks is believed to be driven primarily by their internal stresses \citep{papaloizou_1995}. Gross features of this process can be understood by treating the disk as a geometrically thin structure and characterizing its properties using the vertically integrated (or properly averaged across the disk thickness) variables such as the surface density $\Sigma$ \citep{shakura_1973,lynden-bell_1974}. In this approach the steady accretion disks with the radially constant mass accretion rate $\dot M$ necessarily exhibit purely {\it inward} radial motion of gas, driven by the angular momentum redistribution in the disk due to its internal stresses. In other words, the density-weighted vertical average of the radial velocity $u_R$ is always negative, $\langle u_R \rangle_\rho=\Sigma^{-1}\int \rho u_Rdz<0$.

Situation becomes more complicated once one considers the full three-dimensional disk structure, in particular, the profile of the radial velocity $u_R$ as a function of the vertical coordinate $z$. It was first shown by \citet{Urpin} that when the angular momentum transport in the disk is effected by the effective shear viscosity \citep{shakura_1973}, the radial velocity of the gas is actually {\it positive} at $z=0$, implying a radial {\it ouflow} near the disk midplane. The magnitude of positive $u_R$ steadily decreases with the vertical distance from the midplane, and $u_R$ ultimately changes sign at some altitude of order the disk scale height $H$. As a result, a {\it meridional circulation} pattern sets in the poloidal plane of the disk. Radial gas inflow at high $z$ carries more mass inward than is transported out by the midplane outflow, resulting in a net (vertically integrated) inflow of mass towards the accreting object \citep{Urpin,Jacquet}. Thus, even though $\langle u_R \rangle_\rho=\Sigma^{-1}\int \rho u_Rdz$ is still negative, the radial outflow near the midplane actually transports mass out in 3D. This remarkable analytical prediction was later confirmed by numerical hydrodynamical calculations of \citet{Polka}, \citet{Kley1992}, and \citet{Rozhichka}, which explicitly employed $\alpha$-prescription \citep{shakura_1973} to describe the internal stress.

The meridional circulation, if indeed present in accretion disks, should have very important implications. It may provide a natural way of coupling the inner disk to its outer parts: the material and information from the vicinity of an accreting object could potentially be directly transported in the advective fashion farther out in the disk by the near-midplane outflow. 

To provide some illustrative examples, \citet{Takeuchi} have suggested that meridional circulation can drive the outward transport of dust grains in protoplanetary disks. More recently, the samples collected by the {\it Stardust} mission from comet 81P/Wild 2 revealed the presence of large ($>1\mu$m) grains composed of high-temperature minerals that appear to have formed in the inner regions of the Solar nebula \citep{Brownlee,Zolensky}. Their presence in comets that form at tens of AU indicates that global outward radial transport of dust particles could have operated in the Solar nebula \citep{Ciesla2007,Ciesla2009,Hughes,Jacquet}.

Moon is thought to have formed in a collision of a Mars-size planetary embryo with the proto-Earth \citep{Hartmann}, primarily out of the material derived from the impactor \citep{Canup}. This should have resulted in the different compositions of the Moon and the Earth. However, the isotopic ratios of oxygen derived from lunar {\it Apollo} samples were found to be essentially identical\footnote{Terrestrial O isotopic ratios are quite distinct from e.g. their martian values.} to their terrestrial analogs \citep{Wiechert}. To explain this puzzle \citet{Pahlevan} suggested that equilibration of oxygen isotopes in the Earth-Moon system resulted from the rapid radial mixing between the terrestrial magma ocean and the proto-Lunar disk of vapor-melt debris produced by the impact. \citet{Pahlevan} appealed to diffusive turbulent mixing as the mechanism of compositional equilibration inside the disk. However, meridional circulation can also be an important contributor to this process by directly advecting into the disk and mixing the material from the magma ocean via the midplane outflow.

Weakly magnetized accreting objects (cataclysmic variables, classical nova progenitors, neutron stars in low-mass X-ray binaries, etc.) are expected to develop a {\it boundary layer} between the inner edge of the accretion disk and the stellar surface \citep{Popham1,Popham2,Belyaev1,Belyaev2,Belyaev3}. If the material from the near-surface layers of the accretor can be dredged up (evidence for this exists in e.g. classical novae \citealt{Truran,Gehrz}) into the boundary layer by some internal processes, then the near-midplane outflow could subsequently transport it to the outer parts of the disk. Such elemental pollution might spread the metals with unusual abundances across the disk, affecting its observational appearance (e.g. via spectral signatures).

Given the variety of situations, in which meridional circulation may be important, it is natural to ask how robust this phenomenon is. The majority of its analytical and numerical investigations focused on disks, in which angular momentum transport is accomplished by the conventional $\alpha$-viscosity \citep{shakura_1973}. All of them invariably find $u_R(R,0)>0$ implying the near-midplane outflow. However, in many types of disks the transport is more likely to be effected by the {\it magnetorotational instability} (MRI, \citealt{Balbus}), for which the stress tensor $T_{ij}$ is {\it anisotropic}. The anisotropy of $T_{ij}$ has been shown \citep{Jacquet} to play an important role for the meridional circulation, although the numerical evidence on the issue is rather mixed at this point \citep{Fromang,Flock,Suzuki}, see the discussion in \S \ref{sect:disc}. But even without magnetic fields, global simulations by \citet{Stoll2014,Stoll2016} of the purely hydrodynamic {\it vertical shear instability}  (VSI; \citealt{Urpin1998,Urpin2003}) exhibit radial inflow at the midplane, changing to an outflow at high $z$, contrary to the expectations for disks mediated by the shear stress  \citep{Urpin,Takeuchi}.

In this study we address a different aspect of the problem and explore the effect of {\it thermal properties} of the disk on the meridional circulation, while still confining ourselves to the shear viscosity as the source of the angular momentum transport (which allows us to perform detailed analytical investigation). Existing analytical and numerical work has typically assumed the disk to have locally isothermal vertical structure \citep{Urpin,Fromang,Jacquet,Suzuki}. This may be a reasonable assumption for the externally irradiated protoplanetary disks, but should not generally hold in accretion disks dominated by viscous dissipation. Some earlier numerical studies \citep{Polka,Kley1992,Rozhichka} have attempted to use more refined treatments of the disk thermodynamics. However, they were limited in resolution and did not explore the full range of possible thermodynamic regimes. Our present goal is to provide a thorough analysis of the effects of thermal stratification on the meridional circulation in the disk.  

This work is organized as follows. In \S \ref{sect:theory} we present general theoretical description of meridional circulation in accretion disks. We explore conditions, under which the inflow occurs at the disk midplane in \S \ref{sect:conditions}. In \S \ref{sect:numerics}-\ref{sect:results} we verify our theoretical results with 3D viscous numerical simulations. We discuss our findings in \S \ref{sect:disc}, focusing on applications to real systems in \S \ref{sect:applicability}. Our results are summarized in \S \ref{sect:summ}.


\section{Theoretical considerations}  
\label{sect:theory}


We consider a geometrically thin, axisymmetric disk orbiting in a central potential of a point mass $M_\star$;  we work in cylindrical coordinates $(R,z)$. We are interested in the spatial structure of the radial velocity $u_R(R,z)$, which is non zero because of the internal stress $T_{ij}$ operating in the disk. We are interested only in advective, laminar motions of the fluid and do not consider turbulent diffusion, which is often invoked in studies of particle transport in disks \citep{Takeuchi,Hughes}. Thermal structure of the disk is specified in \S \ref{sect:thermodynamics}.

Equations of the radial and vertical balance, describing the equilibrium disk structure, are
\ba     
\Omega^2R-\frac{1}{\rho}\frac{\partial P}{\partial R} & = & \frac{\partial \Phi}{\partial R},
\label{eq:r}\\
\frac{1}{\rho}\frac{\partial P}{\partial z} & = & -\frac{\partial \Phi}{\partial z},
\label{eq:z}
\ea     
where $\rho$ and $P$ are the gas density and pressure, $\Omega$ is the angular frequency, and $\Phi(R,z)=-GM_\star(R^2+z^2)^{-1/2}$ is the central potential (we neglect the disk self-gravity). {In these equations we also neglected viscous terms resulting from the fluid motions in the meridional plane; we comment on the significance of this simplification later.}

Because of the variation of the central gravity with height and radial pressure support in the disk the angular frequency deviates from its Keplerian value $\Omega_{\rm K}\equiv (GM_\star/R^3)^{1/2}$:
\ba
\Omega(R,z) \approx  \Omega_{\rm K}\left(1-\frac{3}{4}\frac{z^2}{R^2}+\frac{1}{2\Omega_{\rm K}^2R^2}\frac{P}{\rho}\frac{\partial \ln P}{\partial \ln R}\right).
\label{eq:Omega}
\ea
This approximate relation follows from our assumption of a geometrically thin disk, for which $H/R\ll 1$, $z/R\ll 1$, where $H(R,z)\equiv c_s/\Omega_{\rm K}$ is the local disk scaleheight, and $c_s^2(R,z)\equiv P/\rho=k_B T(R,z)/\mu$ is the {\it isothermal} sound speed. Note that the last term in parentheses is a function of both\footnote{In particular, one can show with the aid of equation (\ref{eq:z}) that $\Omega$ is independent of $z$ for a barotropic equation of state.} $R$ and $z$. The deviations from purely Keplerian rotation in equation (\ref{eq:Omega}) are at the $(z/R)^2\ll 1$ (second term in parentheses) and $(H/R)^2\ll 1$ (third term) level.

The $\phi$-component of the equation of motion describing the angular momentum balance can be reduced (under reasonable assumptions regarding the amplitudes of the radial and vertical fluid velocities, \citealt{Takeuchi,Fromang}) to the following expression for the radial velocity $u_R$ of the disk fluid \citep{Jacquet}:
\ba     
u_R(R,z) &=& -\frac{1}{\rho}\left(\frac{\partial l}{\partial R}\right)^{-1}
\nonumber\\
& \times & \left[\frac{1}{R}\frac{\partial }{\partial R}\left(R^2T_{R\phi}\right)+\frac{\partial }{\partial z}\left(RT_{z\phi}\right)\right].
\label{eq:uR}
\ea     
Here $l\equiv \Omega R^2$ is the specific angular momentum of the disk fluid, while $T_{R\phi}$ and $T_{z\phi}$ represent the $R-\phi$ and $z-\phi$ components of the internal stress tensor. This expression is accurate at the $(z/R)^2,(H/R)^2\ll 1$ level; it neglects certain terms proportional to the vertical velocity and possible time variability of the azimuthal velocity \citep{Fromang}. It demonstrates that the radial fluid motion of the 3D disk is determined not only by the horizontal but also by the vertical component of the stress tensor. This turns out to be of crucial importance for our work. On the other hand, multiplying equation (\ref{eq:uR}) by $\rho$ and integrating it over $z$, one finds that $\langle u_R \rangle_\rho$ is determined by the $T_{R\phi}$ behavior only \citep{Jacquet}. 

In real astrophysical disks internal stress can be provided by a variety of mechanisms, with the MRI \citep{Balbus} being one of the important possibilities. However, our present focus is still on standard shear viscosity prescription, for which one has
\ba     
T_{R\phi}=-\rho\nu R\frac{\partial\Omega}{\partial R},~~~~
T_{z\phi}=-\rho\nu R\frac{\partial\Omega}{\partial z},
\label{eq:stress}
\ea     
where $\nu$ is a kinematic viscosity.

Previous studies \citep{Fromang,Jacquet} clearly demonstrated that the possible meridional {\it outflow} is most pronounced at the midplane of the disk, at $z=0$. For that reason, our primary goal will be to determine the factors that affect the {\it midplane} value of the radial velocity, $u_R(R,0)$. To that effect, in our analysis we will often take the limit $z\to 0$ in evaluating different expressions and account for the symmetry of the disk with respect to its midplane. The latter implies, in particular that the first derivatives of various fluid variables with respect to $z$, e.g. $\partial\rho/\partial z$, $\partial\Omega/\partial z$, etc., go to zero as $z\to 0$. 

In Appendix \ref{app:deriv} we show that after some straightforward manipulations the midplane value of the radial velocity can be written as
\ba      
u_R(R,0) &=& \frac{\nu}{R}\left[\frac{15}{2}-3\frac{\partial\ln\left(\rho_0\nu_0 R^3\right)}{\partial R}\right.
\nonumber\\
&-&  \left. \delta_\rho-\frac{P_0}{\Omega_{\rm K}^2\rho_0^2}\frac{\partial\ln P_0}{\partial \ln R} \times \frac{\partial^2\rho}{\partial z^2}\Bigg|_{z\to 0}\right],
\label{eq:uR2}
\ea      
where 
\ba
\delta_\rho\equiv \partial\ln\rho_0/\partial\ln R,
\label{eq:deltarho}
\ea
and subscript $0$ implies the value of a particular variable at the midplane, e.g. $\rho_0(R)\equiv \rho(R,0)$, $P_0(R)\equiv P(R,0)$, etc. Definition (\ref{eq:deltarho}) implies that $\delta_\rho$ is closely related to the density power law exponent $p$ used in previous studies \citep{Takeuchi,Fromang}. The two coincide if $\rho_0(R)\propto R^p$, but the power law density profile is not necessary for our results to be valid. Note that the expression (\ref{eq:uR2}) does not make any assumptions about the vertical or radial dependence of the viscosity, although it does rely on the stress model (\ref{eq:stress}).

The two terms in the first line of the equation (\ref{eq:uR2}) originate from $T_{R\phi}$, while the last two terms are due to the vertical stress $T_{z\phi}$. Since the midplane pressure $P_0$ is a decreasing function of $R$, it is clear that a vertical density profile steeply declining with $z$ (i.e. high value of $-\partial^2\rho/\partial z^2$) should be favorable for suppressing the midplane outflow in the disk and driving gas inflow ($u_R<0$) at all heights.

{Note that equations (\ref{eq:r})-(\ref{eq:z}) neglect contributions from viscous stresses arising from shear in the meridional plane. Such terms were considered in studies of \citet{Kley1992}, \citet{Kluzniak}, and \citet{Regev}, resulting in the  additional $O(\alpha^2)$ relative contribution to the expression (\ref{eq:Omega}) for the angular frequency. Their inclusion would, in turn, lead to the emergence of an additional $O(\alpha^3)$ contribution in the expression (\ref{eq:uR2}) for $u_R$. Given the expectation of $\alpha\ll 1$ in real disks, the omission of the viscous terms should not present a problem\footnote{\citet{Kluzniak} and \citet{Regev} find that inclusion of the viscous terms changes midplane outflow to an inflow only for $\alpha\gtrsim 0.7$, although \citet{Kley1992} suggest a lower value of critical $\alpha\approx 0.06$ for this transition.}. This expectation is subsequently confirmed in \S \ref{sect:results} by the good agreement between our analytical theory (which explicitly neglects viscous contributions to $\Omega(R,z)$) and numerical results.}


\subsection{Disk thermodynamics}  
\label{sect:thermodynamics}


So far our treatment was fully general. To make further progress we need to look in more detail into the disk thermodynamics. We will assume a rather general equation of state (EOS) for the disk fluid in the polytropic form
\ba     
P=e^s\rho^\gamma,~~~s(R,Z)=c_V^{-1}S(R,z),
\label{eq:EOS}
\ea        
where $\gamma$ is the polytropic index and  $s(R,z)$ is the scaled (dimensionless) gas entropy $S(R,z)$; $c_V\equiv (\gamma-1)^{-1}k_B/\mu$ is the specific heat capacity. Note that in this work $S(R,z)$ is specified {\it explicitly}, i.e. we do not attempt to calculate it, e.g. from the energy equation by solving for the radiation transfer in the disk. 

An often used assumption of a locally isothermal disk structure \citep{Takeuchi,Fromang,Jacquet} implies that $P/P_0=\rho/\rho_0=\exp(-z^2/2H_0^2)$, so that
\ba     
s^{\rm iso}(R,z)=s_0(R)+\frac{\gamma-1}{2}\left[\frac{z}{H_0(R)}\right]^2,
\label{eq:Siso}
\ea
where $s_0=\ln\left(P_0/\rho_0^\gamma\right)$. A different limit of the locally isentropic disks considered later in \S \ref{sect:isentropic} has $s(R,z)=s(R)$.

Using the ansatz (\ref{eq:EOS}) we show in Appendix \ref{app:deriv} that the last term inside the brackets in the expression (\ref{eq:uR2}) can be written as
\ba     
\frac{\delta_\rho+\delta_T}{\gamma}\left(1+\frac{P}{\Omega_{\rm K}^2\rho}\frac{\partial^2 s}{\partial z^2}\Bigg|_{z\to 0}\right),
\label{eq:term1}
\ea      
where 
\ba
\delta_T\equiv \partial\ln T_0/\partial\ln R,
\label{eq:deltaT}
\ea
and $T_0(R)\equiv T(R,0)$ is the midplane value of the gas temperature. Definition of $\delta_T$ makes it closely related to the temperature power law index $q$ used in other studies of the meridional circulation in accretion disks \citep{Takeuchi,Fromang}, which assume $T_0(R)\propto R^{q}$.

Next we connect the behavior of the second term in the equation (\ref{eq:uR2}) to the global structure of the disk. We introduce the {\it viscous angular momentum flux} across a given radius $F_J$, which is equal to the viscous torque exerted by the inner disk on the outer disk. The concept of $F_J$ is known to be very useful for describing viscous accretion disks in a variety of situations, especially a steady state \citep{lynden-bell_1974,Rafikov2013,Rafikov2016a}. According to the definition, $F_J$ can be calculated by multiplying $T_{R\phi}$ in equation (\ref{eq:stress}) by $2\pi R^2$ and integrating over $z$. Using equation (\ref{eq:stress}) this implies, to $(H/R)^2$ accuracy, that   \ba     
F_J=-\int\limits_{-\infty}^{\infty}2\pi R^3\rho\nu\frac{\partial\Omega}{\partial R}dz = 3\pi l \int\limits_{-\infty}^{\infty}\rho\nu dz,
\label{eq:FJ}
\ea     
where $l\equiv \Omega_K R^2$ and we have assumed a Keplerian rotation profile.
Also, multiplying equation (\ref{eq:uR}) by $2\pi R\rho$ and integrating over $z$ one finds that the mass accretion rate $\dot M$ (defined to be positive for inflow) is related to $F_J$ as \citep{Rafikov2013}
\ba     
\dot M(R)=-2\pi R \int\limits_{-\infty}^{\infty}\rho u_R dz=\frac{\partial F_J}{\partial l},
\label{eq:Mdot}
\ea  
where the derivative is with respect to the specific angular momentum, so that $\partial/\partial l=2(\Omega_{\rm K} R)^{-1}\partial/\partial R$. 

We now make an additional assumption that the disk structure obeys a certain similarity property, namely that
\ba    
S(R,z)=S\left(R,\frac{z}{H_0(R)}\right),~~~~H_0(R)\equiv H(R,0),
\label{eq:simil}
\ea      
where $H_0=c_s(R,0)/\Omega_{\rm K}$ is the local value of the scaleheight at the disk midplane. This is not a very constraining assumption and its adoption should not limit the applicability of our subsequent results.  

With such entropy behavior it follows from equation (\ref{eq:EOS}) that $\rho$ and $P$ also depend on $z$ only in combination $z/H_0$; this makes it natural to expect that $\nu=\nu(R,z/H_0)$ as well. Then equation (\ref{eq:FJ}) implies that 
\ba    
F_J\propto \rho_0\nu_0 H_0 l\propto \rho_0\nu_0 T_0^{1/2}R^2,
\label{eq:FJrel}
\ea    
allowing us to tackle the second term in equation (\ref{eq:uR2}).

Plugging equations (\ref{eq:term1}) and (\ref{eq:FJrel}) into the expression (\ref{eq:uR2}) we finally find that
\ba      
u_R(R,0) &=& \frac{\nu}{R}\left[\frac{9}{2}-3\left(\delta_F-\frac{\delta_T}{2}\right)\right.
\nonumber\\
&-&  \left. \delta_\rho+\frac{\delta_\rho+\delta_T}{\gamma}\left(1+s^{\prime\prime}_{zz}\right)\right],
\label{eq:uR3}
\ea      
where we introduce the shorthand notation 
\ba
s^{\prime\prime}_{zz}\equiv H_0^2\frac{\partial^2 s}{\partial z^2}\Bigg|_{z\to 0}
\label{eq:szz}
\ea
for the (dimensionless) second derivative of the entropy at the disk midplane, and 
\ba
\delta_F\equiv \partial\ln F_J/\partial\ln R.
\label{eq:deltaF}
\ea
Once again, in equation (\ref{eq:uR3}) the first two terms arise from $T_{R\phi}$, while the terms in the second line originate from $T_{z\phi}$.

Equation (\ref{eq:uR3}) represents the main analytical result of this work. It provides a connection between the amplitude (and direction) of the radial velocity at the disk midplane and the vertical thermal stratification characterized by $s^{\prime\prime}_{zz}$. This link has not been established in previous studies of meridional circulation.

\begin{figure}
\vspace{-20pt}
\centering
\includegraphics[width=0.48\textwidth]{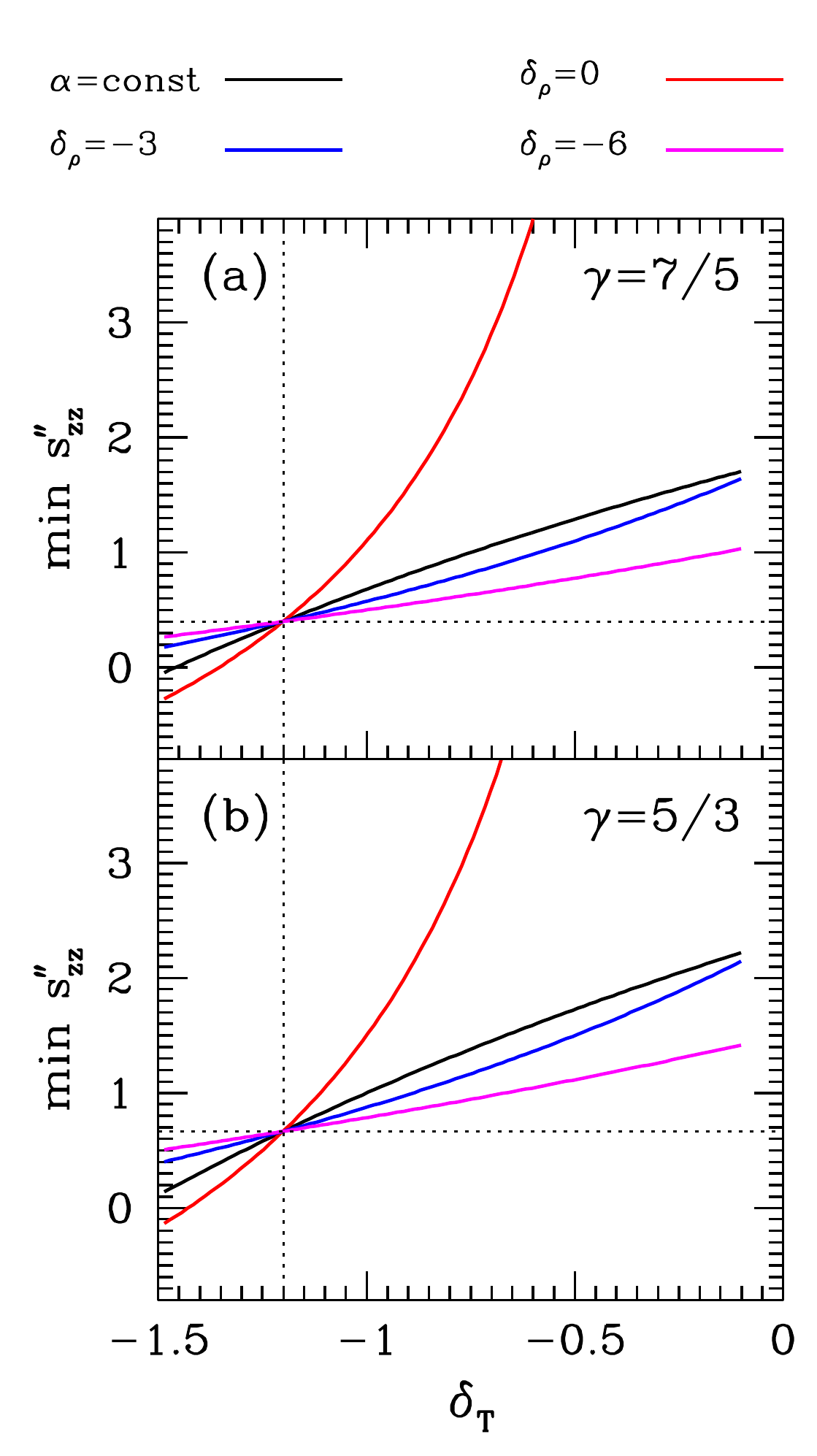}
\caption{Minimum value of $s^{\prime\prime}_{zz}$ --- the second derivative in $z/H_0$ of the scaled entropy $s=S/c_V$ --- necessary for having radial inflow at the disk midplane (i.e. $u_R(R,0)<0$ occurs above a corresponding curve) in a constant $\dot M$ disk ($\delta_F=1/2$), as a function of the midplane temperature slope $\delta_T$ (Eq. [\ref{eq:deltaT}]). Two values of $\gamma$ are considered: (a) $7/5$ and (b) $5/3$. Black curve is a constraint (\ref{eq:crit2}) that applies in a disk with radially constant $\alpha$. Other curves correspond to equation (\ref{eq:crit1}), which does not assume $\alpha=$const, for different values of the midplane density slope $\delta_\rho$ (Eq. [\ref{eq:deltarho}]). Dotted lines are relevant for the locally isothermal disks (\S \ref{sect:isothermal}): horizontal line marks the value of $\partial^2 s/\partial (z/H_0)^2=\gamma-1$ in such a disk, see Eq. (\ref{eq:Siso}). The vertical line marks $\delta_T=-6/5$ --- in locally isothermal disks $u_R(R,0)<0$, i.e. midplane inflow, is possible only for $\delta_T$ below this value.}
\label{fig:constr_theory}
\end{figure}


\section{Conditions for inflow at all $z$}  
\label{sect:conditions}


Equation (\ref{eq:uR3}) allows us to determine the conditions under which the disk will exhibit {\it inflow} at all $z$, including the midplane. Setting $u_R(R,0)<0$ one finds a necessary criterion for this to be the case:
\ba     
s^{\prime\prime}_{zz} && > -\frac{\gamma}{\delta_\rho+\delta_T} 
\nonumber\\
&& \times \left(\frac{9}{2}-3\delta_F+\frac{3\gamma+2}{2\gamma}\delta_T+\frac{1-\gamma}{\gamma}\delta_\rho\right),
\label{eq:crit}
\ea     
where we assumed that $P_0(R)$ decreases with $R$ so that $\delta_\rho+\delta_T<0$. When this inequality is fulfilled meridional circulation is unable to convey information and material from the inner disk to the outer disk --- only {\it inward} propagation is allowed. 

Equation (\ref{eq:crit}) is the most general form of the inflow criterion that does not make any assumptions about the values of $\delta_F$, $\delta_\rho$, and $\delta_T$ --- they can take arbitrary values allowing one to explore meridional circulation even in evolving disks. Moreover, even though it does assume a particular stress model given by the equation (\ref{eq:stress}), it makes only a weak assumption about the actual behavior of the viscosity (that $\nu$ depends on $z$ only in combination $z/H_0$).


\subsection{Inflow criterion for a standard constant $\dot M=$ disk}  
\label{sect:Mdot}


In practice, one is often interested in accretion disks that have reached a steady state. One of the most popular assumptions used in many studies is that of the radially constant mass accretion rate through the disk ($\dot M=$const), with no torque applied at its center. In this case, integrating equation (\ref{eq:Mdot}) one finds $F_J=\dot M l$ \citep{Rafikov2013}, so that $\delta_F=1/2$. This transforms inequality (\ref{eq:crit}) into the following inflow criterion:
\ba     
s^{\prime\prime}_{zz} > -\frac{\gamma}{\delta_\rho+\delta_T} 
\left(3+\frac{3\gamma+2}{2\gamma}\delta_T+\frac{1-\gamma}{\gamma}\delta_\rho\right).
\label{eq:crit1}
\ea     
Note that the constant $\dot M$ assumption does not constrain $\delta_\rho$ or $\delta_T$ because of the freedom in choosing the radial variation of the viscosity $\nu_0$. This constraint is illustrated in Figure \ref{fig:constr_theory} for two values of $\gamma$ ($5/3$ and $7/5$) and three values of $\delta_\rho$. One can see that steeper decay of $\rho_0$ with $R$ (more negative $\delta_\rho$) makes it easier to achieve inflow at the disk midplane, i.e. requires less extreme values of $s^{\prime\prime}_{zz}$.

We can now go one step further and adopt a particular viscosity ansatz, namely the $\alpha$-model of  \citet{shakura_1973}, in which $\nu=\alpha c_s^2/\Omega_{\rm K}$.  If we additionally assume that the effective viscosity parameter $\alpha$ is independent of $R$, then locally $\nu_0\propto R^{\delta_T+3/2}$. Plugging this and $F_J=\dot M l$ into equation (\ref{eq:FJrel}) one obtains that 
\ba     
\delta_\rho=-3-\frac{3}{2}\delta_T.
\label{eq:drho_alpha}
\ea    
This results in yet another version of the inflow criterion for $\alpha$ disks, which depends only on the radial temperature profile (i.e. $\delta_T$):
\ba     
&& s^{\prime\prime}_{zz} > \frac{6(2\gamma-1)+(6\gamma-1)\delta_T}{6+\delta_T}.
\label{eq:crit2}
\ea     
This constraint is also shown in Figure \ref{fig:constr_theory}.


\subsection{Effects of thermal structure of the disk}  
\label{sect:thermal}


Next we assess how the different assumptions about the vertical thermal structure of the constant $\dot M$ disk affect the possibility of the inflow at its midplane.


\subsubsection{Locally isentropic disk}  
\label{sect:isentropic}


The simplest thermodynamic assumption is that of a {\it locally isentropic} disk, in which entropy (and $s$) does not depend on $z$. Such disks naturally have temperature decreasing with height. The isentropic vertical stratification may arise e.g. if the disk is convectively unstable.

In this case the left hand side of equation (\ref{eq:crit1}) becomes zero and the condition of inflow turns into a constraint on the radial behavior of $\rho_0(R)$ and $T_0(R)$ in the form
\ba    
\delta_\rho>\frac{6\gamma+(3\gamma+2)\delta_T}{2(\gamma-1)}.
\label{eq:deltarhoconstr}
\ea     
For protoplanetary disks with $\gamma=7/2$ this becomes $\delta_\rho>(42+31\delta_T)/4$. In hotter accretion disks with $\gamma=5/3$ the constraint is $\delta_\rho>3(10+7\delta_T)/4$.

It is clear that even if $T_0(R)$ decays with radius as rapidly as $R^{-1}$ (i.e. $\delta_T=-1$) the inflow at the disk midplane would require midplane density {\it increasing outwards}. This is a pretty unusual arrangement, which makes midplane inflow essentially impossible in locally isentropic disks, including the disks which are convectively unstable.


\subsubsection{Locally isothermal disk}  
\label{sect:isothermal}


A popular assumption of the locally isothermal disk structure ($\partial T(R,z)/\partial z=0$) turns the left hand side of the inflow criterion into $\gamma-1$, see equation (\ref{eq:Siso}). Then it is easy to see from the inequality (\ref{eq:crit1}) that the inflow at all heights is possible only if $\delta_T<-6/5$. This result is completely independent of either $\delta_\rho$ or $\gamma$, thus it should hold for an arbitrary radial profile of the midplane density $\rho_0(R)$. 

Finding conditions in which a disk might have $T_0(R)$ decaying with radius faster than $R^{-6/5}$ is not easy. Based on this we can conclude that the locally isothermal constant $\dot M$ disks are predetermined to naturally exhibit an outflow at the midplane, in agreement with a number of previous studies \citep{Urpin,Jacquet}. 

Summarizing the results of this and previous subsection (\S \ref{sect:isentropic}), we conclude that the inflow at $z=0$ requires the disk to have {\it temperature increasing with height}, starting from the midplane. Such disks naturally have entropy rising more steeply with $z$ than in the isothermal case and may satisfy the conditions (\ref{eq:crit1}) and (\ref{eq:crit2}) without making unrealistic assumptions about the behavior of $\rho_0(R)$ and $T_0(R)$.


\subsection{Inflow criterion for $F_J=$ const disk}  
\label{sect:FJ}


\begin{figure}
\centering
\vspace{-50pt}
\includegraphics[width=0.5\textwidth,scale=0.3]{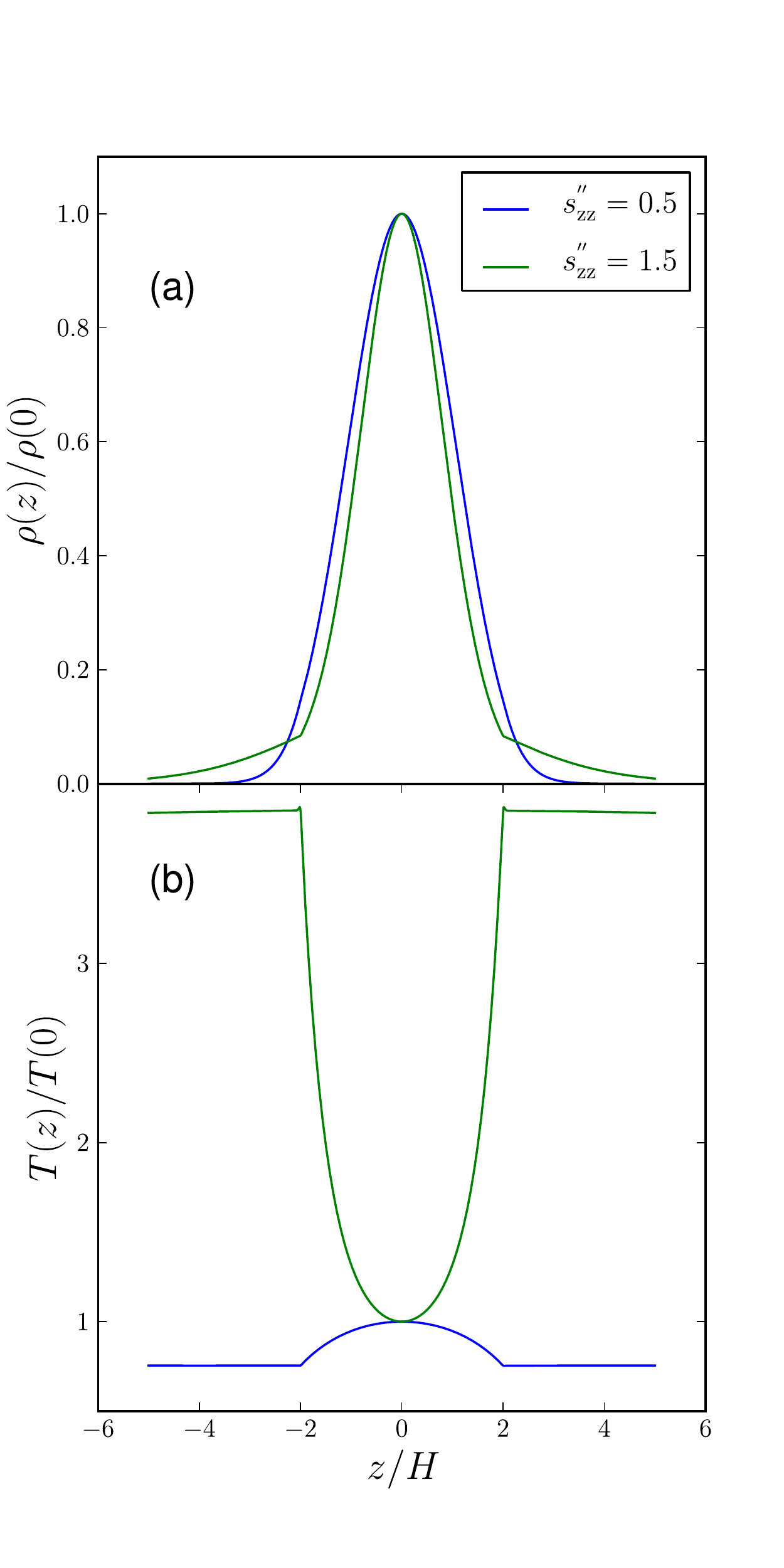}
\caption{Vertical profiles of the density (a) and temperature (b) in a constant $\dot M$ disk with the radial temperature profile such that $\delta_T=-1$ (see Eq. [\ref{eq:deltaT}]) and two vertical entropy profiles characterized by  $s^{\prime\prime}_{zz}=0.5$ (blue) or $1.5$ (green), see the definition (\ref{eq:szz}). All variables are normalized by their midplane values. Depending on the value of $s^{\prime\prime}_{zz}$, our thermodynamic anzatz (\ref{eq:EOS}) can result in either vertically decaying (for low $s^{\prime\prime}_{zz}$) or rising (for high $s^{\prime\prime}_{zz}$) temperature profiles near the midplane.}
\label{fig:initial}
\end{figure}

Standard constant-$\dot M$ disk with zero central torque represents just one particular example of a steady disk. A more general time-independent structure of the disk is described by $F_J(R)=\dot Ml+F_J(0)$ \citep{Rafikov2013,Rafikov2016a}, where $F_J(0)$ represents the torque applied at the disk center (which is naturally absent in a standard constant-$\dot M$ case). 

A particularly interesting case to explore is that of a disk with {\it no mass accretion} at the center, in which the gas inflow is fully suppressed by a strong central torque \citep{lynden-bell_1974,Pringle,Rafikov2016b}. Such "dead" disks were first studied by \citet{SyunShak} in the context of accretion by the magnetized neutron stars in the propeller regime \citep{Illarionov}. The global structure of such disks is characterized by $F_J(R)=$const, implying that $\delta_F=0$.

Equation (\ref{eq:uR3}) demonstrates that reaching inflow at the midplane of such a disk is {\it more difficult} than in the standard constant-$\dot M$ disk. Indeed, the inflow criterion (\ref{eq:crit1}) remains essentially the same, however the first term inside the parentheses changes from 3 to 9/2. As a result, a higher value of $s^{\prime\prime}_{zz}$ is needed to guarantee inflow at all heights (i.e. $u_R(R,0)<0$) in the constant-$F_J$ disk, as compared to the standard constant $\dot M$ disk (\S \ref{sect:Mdot}). Inequality (\ref{eq:crit2}) gets modified in a similar fashion.

Moreover, repeating the calculations of \S \ref{sect:isothermal} one finds that the inflow at $z=0$ in a locally isothermal, constant-$F_J$ disk would require $\delta_T<-9/5$. Such steeply declining radial profiles of temperature are unlikely in real disks. 

\begin{figure*}
\hspace*{-2cm}
\includegraphics[width=1.2\textwidth]{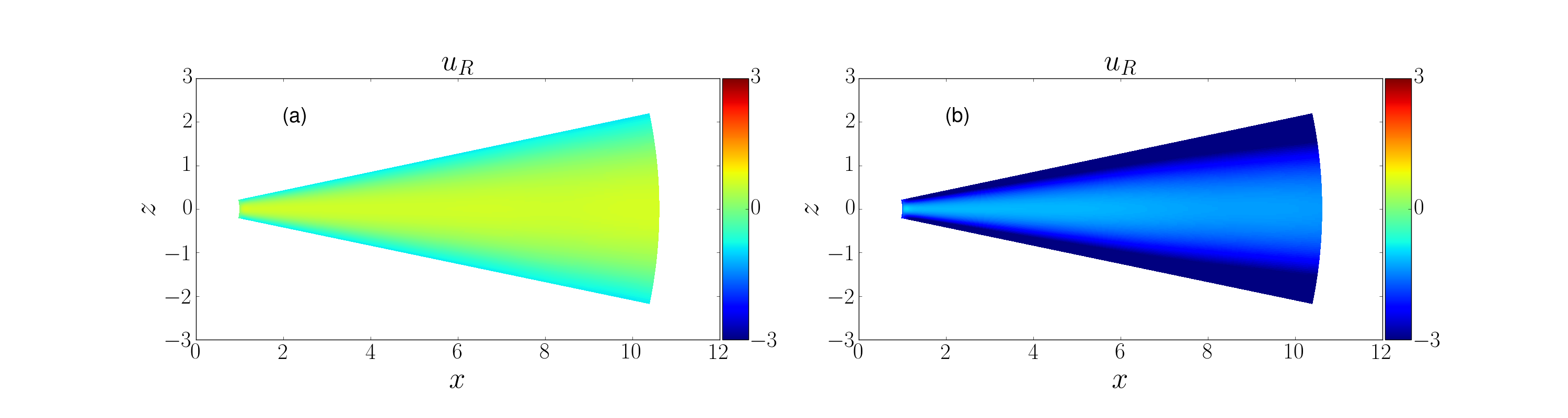}
\caption{Maps of radial velocity $u_R(R,z)$, averaged over azimuth and normalized by $\nu/R$, for $T(R)$ profile with $\delta_T=-1$, constant accretion rate $\dot{M}$ (i.e. $\delta_F=1/2$) and two values of the second derivative of entropy with height, $s^{\prime\prime}_{zz}$:  (a) $s^{\prime\prime}_{zz} = 0.5$ (entropy slowly increasing with $z$), which exhibits outflow at the midplane and inflow at high altitudes; (b) $s^{\prime\prime}_{zz} = 1.5$ (entropy rapidly growing with $z$), which shows gas inflow at all altitudes. }
\label{fig:Urmap}
\end{figure*}


\section{Numerical setup}  
\label{sect:numerics}


In order to confirm the analytical result (\ref{eq:uR3}), we performed viscous hydrodynamical simulations using new Godunov code Athena++ code (Stone et. al. 2016, in preparation). Compared with its predecessor Athena \citep{athena}, Athena++ is highly optimized and uses flexible grid structures, significantly facilitating global numerical simulations. In this work we perform our simulations in spherical $r,\theta,\phi$ coordinates, using uniform grid in $\log r$ and constraining ourselves to small range in $\phi$. We use the domain $[1,10]\times[\pi/2-0.5,\pi/2+0.5]\times[0,0.1]$ in these coordinates with numerical resolution of $1024\times 512\times 16$ cells. We verified that simulations in the full $\phi$ range $[0,2\pi]$ lead to same results. We also verified that our results are converged with regard to numerical resolution. 

For $r$ and $\theta$ boundaries, we use the so-called "do-nothing" boundary condition, i.e. all fluid variables are fixed at their initial values. In $\phi$-direction, we impose a periodic boundary condition. The Mach number of the orbital flow at the inner radius is $10$ in all our simulations. The simulation time is always measured in units of $\Omega_K^{-1}$, where $\Omega_K$ is the Keplerian angular velocity at the inner boundary.

In order for the disk structure to remain unchanged and maintain its initially prescribed entropy profile during the simulation, we use the optically thin cooling function:
\begin{equation}
\Lambda = - \rho \frac{T-T_0}{\tau},
\end{equation}
where $T$ is the gas temperature, $T_0(R,z)$ is the initial temperature profile described in \S \ref{sect:initial}, $\tau=0.5\Omega_K^{-1}$ is the cooling time. This cooling prescription ensures that viscous heating does not significantly influence our initial disk configuration, and that our simulations reach steady state. 

\begin{figure*}
\hspace*{-2cm}
\includegraphics[width=1.2\textwidth]{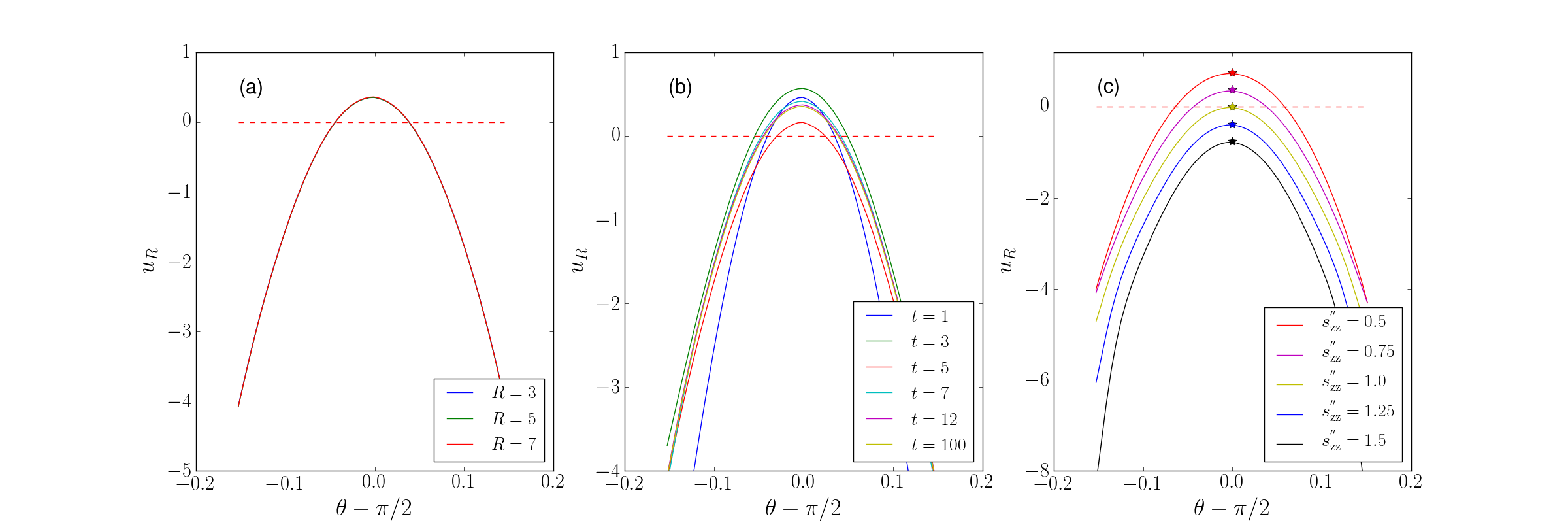}
\caption{Slices of radial velocity $u_R(R,z)$, averaged over azimuth and normalized by $\nu/R$ for $\delta_T=-1$ and constant accretion rate $\dot{M}$. (a): Slices of normalized $u_R$ at different radii from the center at $t=100$ for $s^{\prime\prime}_{zz}=0.75$, demonstrating the self-similar character of the meridional circulation (all three curves fall on top of each other);
(b) Normalized $u_R$ profiles at $R=5$ for $s^{\prime\prime}_{zz}=0.75$ at different moments of time, showing the convergence of $u_R$ to a steady-state; (c): Same at $R=5$ in steady state for different values of $s^{\prime\prime}_{zz}$, characterizing the falloff of the vertical entropy profile. Theoretical predictions for $u_R$ at $z=0$ (equation \ref{eq:uR3}) are shown as stars. One can see both the excellent agreement with theory and the fact that steeply increasing entropy profiles result in gas inflow at all altitudes (see \S \ref{sect:results}).}
\label{fig:cuts}
\end{figure*}


\subsection{Initial entropy and density profiles}  
\label{sect:initial}


For simplicity, we assume the power-law behavior of the density and temperature at the mid-plane $\rho_0(R)\propto R^{\delta_\rho}$, $T_0(R)\propto R^{\delta_T}$. Our simulations adopt a setup typical for standard constant $\dot M$ disks (i.e. we do not simulate disks considered in \S \ref{sect:FJ}), making different assumptions about the radial profile of $\alpha$. Close to the mid-plane, we consider the following entropy profile
\begin{equation}
s(R,z)=s_0(R)+\frac{s^{\prime\prime}_{zz}}{2}\left[\frac{z}{H_0(R)}\right]^2,
\label{eq:sused}
\end{equation}
such that the definition (\ref{eq:szz}) holds true; also $s_0(R)=\ln\left(P_0/\rho_0^\gamma\right)$. In order to prevent large gradients of fluid variables from developing high above the mid-plane, we modify this entropy behavior in such a way that the disk becomes locally isothermal at $|z|>2H_0$. 

To compute the vertical disk structure corresponding to this entropy behavior, we numerically integrate the equation of hydrostatic equilibrium  (\ref{eq:z}). The azimuthal component of the gas velocity is then computed using equation (\ref{eq:r}). We show disk profiles for $\delta_T=-1$, constant $\dot M$ and two values of $s^{\prime\prime}_{zz}$ in Figure \ref{fig:initial}. It shows that for high values of the entropy derivative $s^{\prime\prime}_{zz}$ the disk naturally develops temperature profile rising with height.  As we show below, in this case one finds gas inflow at the disk midplane. On the contrary, low values of  $s^{\prime\prime}_{zz}<\gamma-1$ result in $T$ dropping with $z$ (by design, at high altitudes our $T$ profiles always converge to isothermal).

To check our analytical prediction (\ref{eq:uR3}), we run two sets of simulations. In a first set, we adopt a radially constant $\alpha$ profile with $\alpha=0.01$, so that equations (\ref{eq:drho_alpha}) and (\ref{eq:crit2}) hold. In a second set, we assume that $\alpha \propto R$, with $\alpha=0.01$ at $R=1$. In this case, the constant $\dot M$ assumption leads to a different relation between the power-law indices of density and temperature profiles:
\ba     
\delta_\rho=-4-\frac{3}{2}\delta_T.
\label{eq:drho_alpha1}
\ea 
Plugging this $\delta_\rho$ in equation (\ref{eq:crit1}), we obtain a new criterion for the inflow at all $z$, which replaces equation (\ref{eq:crit2}):
\begin{equation}
s^{\prime\prime}_{zz} > \frac{2(7\gamma-4)+(6\gamma-1)\delta_T}{8+\delta_T}.  
\label{eq:crit3}
\end{equation}
This constraint is again a function of the radial temperature profile only.

For both prescriptions for $\alpha$-viscosity, we perform simulations for several sets of different values of $s^{\prime\prime}_{zz}$ and $\delta_T$.


\section{Simulation results}  
\label{sect:results}


\begin{figure}
\centering
\vspace{-50pt}
\includegraphics[width=0.5\textwidth]{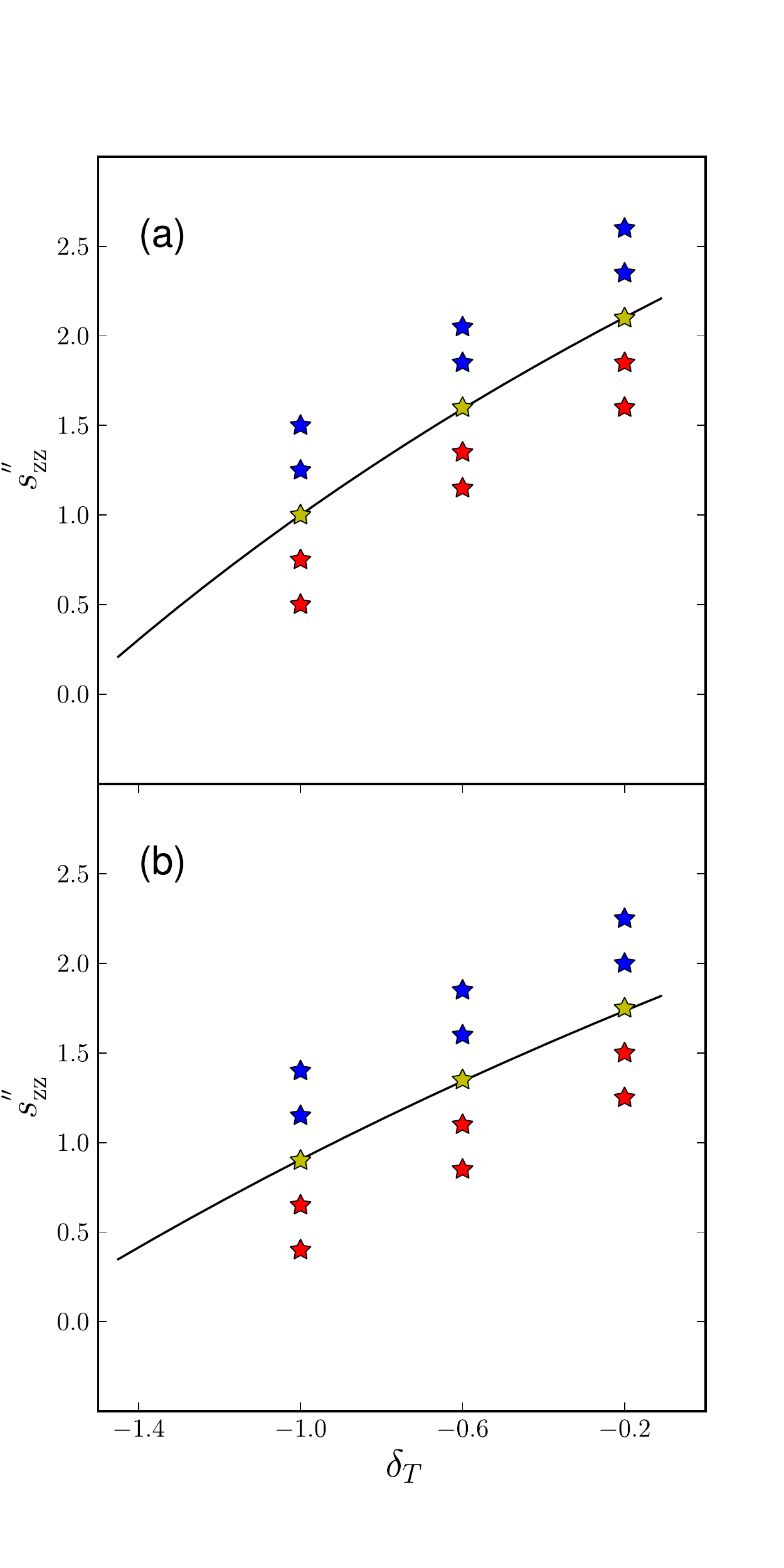}
\caption{Summary of simulation results for different values of $\delta_T$ and $s^{\prime\prime}_{zz}$, and different radial behaviors of viscous $\alpha$: (a): $\alpha=$const,  (b) $\alpha\propto R$. Red, yellow and blue stars represent simulations showing outflow, $u_R(R,0)\approx 0$, and inflow at the midplane, correspondingly. Solid curves represent analytical predictions given by equations (a) (\ref{eq:crit2}) and (b) (\ref{eq:crit3}). One can see excellent agreement between theory and simulations (e.g. yellow stars fall on top of the curves).}
\vspace{-10pt}
\label{fig:res}
\end{figure}

We start by presenting the results of simulations for $\alpha(R)=$const case. In Figure \ref{fig:Urmap} we show the steady state distribution of the radial velocity $u_R$ in the $R-z$ plane for $T(R)$ profile with $\delta_T=-1$ and two values of $s^{\prime\prime}_{zz}=0.5$ and 1.5. The two disks have identical radial profiles of all fluid variables in the midplane, but different vertical profiles of $\rho$ and $T$. Clearly, this leads to an important difference: while $s^{\prime\prime}_{zz}=0.5$ case shows outflow of gas in the midplane and inflow at high altitudes, the $s^{\prime\prime}_{zz}=1.5$ run shows {\it inflow at all altitudes}, in agreement with analytical expectations (as shown in Figure \ref{fig:initial}, in this case $T$ increases with height). 

In Figure \ref{fig:cuts}a we show meridional profiles of the radial velocity, normalized to $\nu/R$, at different radii for $s^{\prime\prime}_{zz}=0.75$. Given our thermodynamic prescription (\ref{eq:sused}), we expect and find to hold in simulations the following parabolic profile of the radial velocity: $u_R(R,z) \propto \left[1+C\times(z/H_0)^2\right]$, $C<0$ (c.f. \citealt{Jacquet}). In the case of $\delta_T=-1$, when $H_0=R$ this behavior results in the meridional profile of $u_R$ normalized by $\nu/R$, which is independent of $R$. This self-similarity is indeed observed in our simulations, as demonstrated by the overlapping curves in Figure \ref{fig:cuts}a. 

In Figure \ref{fig:cuts}b we demonstrate the convergence of the $u_R$ profile to a steady state. One can see that convergence happens within several tens  $\Omega_K^{-1}$, which is much faster than the local viscous time scale in the disk, in agreement with earlier studies \citep{Polka, Kley1992, Rozhichka}.  

In Figure \ref{fig:cuts}c we show how the steady-state vertical profiles of $u_R$ (at fixed $R$ and $\delta_T$) vary as we change the value of $s^{\prime\prime}_{zz}$. One can see that increasing the rate at which gas entropy grows with height uniformly shifts the $u(R,z)$ profile down. Stars at $z=0$ show theoretical prediction  (\ref{eq:uR3}) for the value of $s^{\prime\prime}_{zz}$ corresponding to each curve. Our simulation results obviously match theory very well. 

To test the inflow criterion (\ref{eq:crit2}) we run a set of such simulations for different values of $\delta_T=-1.5,-1,-0.5$. In each run, we measure the radial component of the gas velocity in the midplane $u_R(R,0)$ and record its sign. The outcome of these runs is shown in Figure \ref{fig:res}a, where we mark solutions with $u_R(R,0)<0$ as blue stars, with $u_R(R,0)>0$ as red, and simulations with $u_R(R,0) \approx 0$ as yellow. Analytical inflow criterion (\ref{eq:crit2}) is shown as a solid line. We find that for each inspected value of $\delta_T$, there is a critical value of the vertical entropy gradient, $s^{\prime\prime}_{zz}$, above which the disk exhibits the inflow at all $z$. This Figure demonstrates that this critical value is in very good agreement with the criterion (\ref{eq:crit2}).

We also repeated the same set of simulations for a different viscosity behavior $\alpha \propto R$. Their results, together with the analytical criterion (\ref{eq:crit3}), are summarized in Figure \ref{fig:res}b. As in the case of $\alpha=$const, we find very good agreement between theory and simulations.


\section{Discussion}  
\label{sect:disc}


Our study illuminates the direct effect of the disk thermodynamics on the pattern of meridional circulation. Our finding that in viscous disks the entropy rapidly rising with height may reverse the radial outflow at the midplane, resulting in inflow at all heights, is a new result. It was not found in previous studies focused primarily on the locally isothermal disks for the following reason. 

According to equation (\ref{eq:uR}), meridional circulation is driven by both radial and vertical components of the stress tensor. Starting with the former, large negative radial gradient of density at the midplane makes the net contribution of $T_{r\phi}$ to $u_R(R,0)$ positive, promoting a midplane outflow \citep{Takeuchi}. On the other hand, at high altitudes the density gradient with respect to $R$ changes sign, so that $T_{r\phi}$ drives inflow at high $z$. This leads to the characteristic parabolic shape of the vertical $u_R$ profiles, changing sign at some intermediate altitude.

Vertical stress is a bit more subtle. As the second line of equation (\ref{eq:deltarho}) shows, it contains a contribution (first term) proportional to the radial density gradient that drives an outflow. But the second term proportional to $\partial^2\rho/\partial z^2<0$ promotes inflow in non-pathological cases when $P_0$ decreases with $R$. It arises because the gas at higher $z$ rotates slower than the midplane (see equation [\ref{eq:Omega}]) and viscously removes angular momentum from the underlying layers, forcing them to flow inwards. The efficiency of this coupling near the midplane is regulated primarily by the effect of the vertical variation of $\rho$ on $\Omega$: density slowly decaying with $z$ results in lower $|\partial^2\Omega/\partial z^2|$, see equation (\ref{eq:Omegazz2}). This results in less efficient angular momentum removal from the midplane layers, which cannot oppose the outflow tendency at the midplane. Such situation holds in locally isothermal disks, whose gaussian density profile does not decay rapidly enough to prevent an outflow near the midplane. 

However, when the entropy increases with height more rapidly, the density falls off with $z$ more steeply, resulting in larger vertical shear. This leads to a stronger angular momentum loss from the midplane to the upper layers, making it possible to suppress the positive contribution of the $T_{r\phi}$ to $u_R(R,0)$. As a result, gas can flow in at all heights. 

{Good agreement found between our analytical theory and the results of viscous hydrodynamical simulations, evident in Figures \ref{fig:cuts}c \& \ref{fig:res}, suggests that our results can be used for code testing purposes. They can also be called upon for designing hydrodynamical viscous (not MHD) simulations, which do not exhibit an outflow at the midplane. This setup can be useful e.g. in the long-term studies of the boundary layers of accretion disks. Such simulations are numerically expensive to be performed with full MHD in 3D, calling for a viscous hydro approach instead. But one might also want to suppress the outward fluid motions anywhere in the disk to exclude the effect of the boundary layer on the boundary conditions at the outer edge of the simulation domain. Our results demonstrate that this can be naturally achieved just by making the vertical profile of entropy in the disk to rapidly increase with height.}

{All our results hold for a particular stress model  --- shear stress --- represented by equations (\ref{eq:stress}). In practice, angular momentum transport in hot, well ionized disks is expected to be effected by the MRI \citep{Balbus}. In the weakly ionized regions of the protoplanetary disks, where the MRI may be inactive, the VSI \citep{Urpin1998,Urpin2003} has been suggested to drive angular momentum transport. In both cases the stress behavior may be different from that given by equations (\ref{eq:stress}); we discuss this possibility next.}


\subsection{Circulation in the MRI-dominated disks}  
\label{sect:MRI}

Significant effort has been invested in quantifying the $T_{r\phi}$ behavior for the MRI, which determines $\langle u_R \rangle_\rho$ and, thus, sets the accretion rate $\dot M$. As found by many numerical studies for locally isothermal disks, $T_{r \phi}$ is nearly constant at low altitudes and falls off sharply in the corona, where plasma $\beta$ parameter is of the order of unity, and MRI is quenched \citep{Fromang,Flock,Bai13}. Considerably less is known about the behavior of $T_{z\phi}$, although some numerical results were presented in \citet{Fromang}. 

Given that the MRI stress behavior differs from that of the shear stress, it is not surprising that in their isothermal simulations with no vertical field \citet{Fromang} found a very different circulation pattern, namely an {\it outflow at all altitudes}, so that $u_R$ does not change sign. \citet{Jacquet} explained this result by adopting a phenomenological prescription for the MRI stress behavior similar to the one found by \citet{Fromang}. The major difference of this prescription, when compared to the anzatz (\ref{eq:stress}), is that neither $T_{r \phi}$ nor $T_{z \phi}$ was assumed\footnote{We note, however, that the vertical profile of $T_{z\phi}$ found by \citet{Fromang} is not incompatible with the prescription (\ref{eq:stress}).} by \citet{Jacquet} to scale with density near the midplane. As a result, the logic used above to explain high-altutude inflow in isothermal disks with shear stress fails, and it becomes possible to have pure outflow solutions even with locally isothermal EOS. 

It should, however, be remembered that the picture of the meridional circulation in MRI-active disks is still far from complete. For example, in their simulations with very similar setup (starting with weak toroidal field) \citet{Flock} found a behavior of the $u_R(R,z)$ different from \citet{Fromang}: inflow at the midplane, and outflow at high $z$, i.e. $u_R(R,z)$ profile does change sign. 

{In a somewhat different setup --- starting with a net vertical field --- \citet{Suzuki} observed weak gas {\it outflow} at the midplane, with clear inflow higher up, i.e. the circulation pattern typical for disks with shear viscosity. This finding has been recently confirmed by the non-ideal MHD simulations of \citet{Bethune} and ideal MHD simulations of \citet{Zhu}: in both studies the midplane outflow changing to an inflow at high altitudes was observed. Even though this pattern of circulation agrees with the prediction of the conventional viscous disk theory \citep{Urpin}, the true reason for such behavior in these simulations is likely quite complicated. In particular, \citet{Zhu} find that magnetic stresses play more important role that the thermal pressure gradients in determining the vertical profile of $\Omega$ near the midplane. Clearly, this feature cannot be captured in the framework of our purely hydrodynamical model. }

{The differences between the aforementioned MRI studies strongly suggest that further global stratified MRI simulations with different initial field geometries and varied thermodynamics are needed to clarify both the behavior of $u_R(R,z)$ and the vertical variation of the stress tensor (including $T_{z\phi}$!), as well as to establish the connection between them. }


\subsection{Circulation in the VSI-dominated disks}  
\label{sect:MRI}

{\citet{Stoll2014,Stoll2016} found that the weak transport associated with the VSI gives rise to a near-midplane {\it inflow} of gas, switching to an outflow at high altitudes. Even though VSI is a purely hydrodynamical instability, this behavior is clearly different from that expected in disks governed by the shear stress with isotropic viscosity (i.e. when the radial and vertical stress components $T_{r\phi}$ and $T_{z\phi}$ are characterized by the same value of $\nu$), regardless of their entropy profile, see Figure \ref{fig:cuts}c. }

{As shown in the recent work of \citet{Stoll2017}, this behavior is caused by a strong anisotropy of the stress tensor in the VSI-dominated disks: $T_{z\phi}$ was found to exceed $T_{R\phi}$ by a factor of several hundred. Moreover, the results of \citet{Stoll2017} suggest that near the midplane the vertical variation of $T_{z\phi}$ can be reasonably well approximated by the shear stress prescription (\ref{eq:stress}). The radial stress $T_{R\phi}$ certainly does not follow the scaling $T_{R\phi}\propto \rho$ postulated in equations (\ref{eq:stress}), as \citet{Stoll2014} and \citet{Stoll2017} find $T_{R\phi}$ to {\it increase} with $z$ near the midplane, switching to a decay only at high altitudes. However, given the negligible role played by $T_{R\phi}$ in the VSI-dominated disks, we can nevertheless use our results to understand the meridional circulation pattern observed in simulations of such disks.}

{Indeed, let us look at the equation (\ref{eq:uR3}), in which we will drop the terms in the first line, as they result from the negligible $T_{R\phi}$. We also set $s^{\prime\prime}_{zz}=\gamma-1$ as appropriate for the vertically isothermal disk structure often adopted in the VSI simulations \citep{Stoll2017}. Then the terms in the second line of equation (\ref{eq:uR3}), coming from the $T_{z\phi}$, naturally result in $u(R,0)=\nu_z\delta_T/R$, where $\nu_z$ is the value of the kinematic viscosity coefficient characterizing the vertical behavior of the $T_{z\phi}$ as found in Kley et al. Given that the disk temperature decreases with $R$ (i.e. $\delta_T<0$), one finds that $u(R,0)<0$ (meaning midplane inflow) in the VSI-mediated disks dominated by the vertical stress. Thus, our analytical calculations provide a natural explanation for the meridional circulation pattern found in simulations of such disks.}

Despite all the complications related to stress anisotropy, based on our results, we would still expect thermal stratification to have an important effect on the meridional circulation in 3D disks even if the angular momentum transport is mediated by mechanism other than the shear viscosity.


\subsection{Applicability to real astrophysical systems}  
\label{sect:applicability}

Our study shows that whenever the angular momentum transport in the disk is effected by shear viscosity, the radial inflow at all altitudes above the disk midplane necessarily requires both temperature and entropy to rapidly increase with $|z|$. In light of this result, it is natural to ask, in which astrophysical systems such conditions could hold. 

Protoplanetary disks, heated predominantly by stellar irradiation, have roughly isothermal vertical temperature profile within several scaleheights above the midplane \citep{Chiang}. This expectation motivated the adoption of the isothermal thermodynamic setup in many previous studies of meridional circulation \citep{Urpin,Ciesla2007,Jacquet}. Presence of the superheated dust layer high above the midplane should eventually result in temperature rise at some $z$ \citep{Chiang}, but this is insufficient to revert the radial outflow in the near-midplane part of the disk. Thus, passive protoplanetary disks should feature a near-midplane outflow, if their transport is governed by equations (\ref{eq:stress}).

Disks heated predominantly by viscous dissipation and accreting at high $\dot M$ are likely to be optically thick; their $T(R,z)$ profile inevitably exhibits temperature dropping with height. Once again, our study shows that such disks should exhibit radial outflow near the midplane. Disks with isentropic stratification ($s^{\prime\prime}_{zz}=0$) may feature the fastest radial outflow at $z=0$, since entropy dropping with $z$ will likely drive efficient convection enforcing vertically homogeneous entropy.

However, viscously heated disks accreting at low $\dot M$ should be optically thin to their own emission \citep{Narayan}. If the specific viscous dissipation rate increases with height (which is expected, for example, if heating is produced by the MRI \citep{Miller2000,Hirose06}), then gas temperature will increase with $z$ at all altitudes. This can potentially provide the conditions favorable for the radial inflow at all heights, as we showed in this work. Thus, optically thin, viscously heated accretion disks present the best setting for suppressing the radial outflow at all altitudes.


\section{Summary}  
\label{sect:summ}


We explored meridional circulation in accretion disks with shear viscosity and varied thermodynamics. While previous studies of this problem focused on the  vertically isothermal disks, finding radial gas outflow at the midplane and inflow at high latitudes, we demonstrate that different assumptions about thermal stratification can change this pattern. We show that the direction of the flow at the midplane is intimately connected to the behavior of the vertical stress $T_{z\phi}$. Vertical density profiles steeply falling off with height induce significant vertical shear, which can make the gas at the disk midplane to flow inward. We derive analytical criterion relating the direction of the midplane flow to the thermal stratification in the disk and show that the radial inflow at all heights (without change of sign of $u_R$) naturally sets in disks with steeply growing vertical profiles of entropy and temperature. Such conditions can be naturally realized in optically thin disks heated primarily by viscous dissipation. Although our findings rely on the assumption of shear viscosity, we also comment on other mechanisms of angular momentum transport. Our results can also be used for code testing and designing simulations with the prescribed pattern of meridional circulation.  

\acknowledgements

We are grateful to Willy Kley for useful discussions. A.A.P. is supported by Porter Ogden Jacobus Fellowship, awarded by the graduate school of Princeton University. Financial support for this study has been provided by the NSF via grant AST-1515763, NASA via grants 14-ATP14-0059, and The Ambrose Monell Foundation. Simulations presented in this article used computational resources supported by the PICSciE-OIT TIGRESS High Performance Computing Center and by NSF through an XSEDE computational time allocation TG-AST160008 on TACC Stampede and Ranch.

\appendix


\section{Derivation of $u_R$.}  
\label{app:deriv}


Plugging the expressions (\ref{eq:stress}) into the equation (\ref{eq:uR}) and using the disk symmetry with respect to $z=0$, one finds that the radial velocity at the disk midplane is
\ba     
u_R(R,0) &=& \frac{2\nu R}{\Omega_K}\left[\frac{\partial^2\Omega}{\partial R^2}+\frac{\partial\Omega}{\partial R}\times\frac{\partial\ln\left(\rho\nu R^3\right)}{\partial R}\Bigg|_{z\to 0}+\frac{\partial^2\Omega}{\partial z^2}\Bigg|_{z\to 0}\right].
\label{eq:uR1}
\ea      
Here we approximated $l=\Omega_{\rm K}R^2$, which is accurate at the $(H/R)^2$ level. To the same degree of accuracy we can replace $\Omega$ with $\Omega_{\rm K}$ in all terms with the radial derivatives of $\Omega$. However, the last term in the brackets must be treated more carefully. Using the relation (\ref{eq:Omega}) one finds
\ba     
\frac{\partial^2\Omega}{\partial z^2}\Bigg|_{z\to 0} &\approx & \frac{\Omega_{\rm K}}{2R^2}\left[-3 +\frac{1}{\Omega_{\rm K}^2}\frac{\partial^2}{\partial z^2}\left(\frac{1}{\rho}\frac{\partial P}{\partial \ln R}\right)\Bigg|_{z\to 0}\right]
\label{eq:Omegazz1}\\
&\approx & -\frac{\Omega_{\rm K}}{2R^2}\left[\delta_\rho +\frac{P}{\Omega_{\rm K}^2\rho^2}\frac{\partial\ln P}{\partial \ln R} \frac{\partial^2\rho}{\partial z^2}\Bigg|_{z\to 0}\right],
\label{eq:Omegazz2}
\ea      
where $\delta_\rho$ is defined by equation (\ref{eq:deltarho}). In going from (\ref{eq:Omegazz1}) to (\ref{eq:Omegazz2}) we used equation (\ref{eq:z}) and disk symmetry with respect to its midplane. Plugging result (\ref{eq:Omegazz2}) into the equation (\ref{eq:uR1}) one arrives at the expression (\ref{eq:uR2}).

Further progress involves the knowledge of the EOS of the disk fluid. Plugging the anzatz (\ref{eq:EOS}) into the equation (\ref{eq:z}), taking a derivative of both sides with respect to $z$, and using the symmetry property at $z=0$ one finds
\ba    
\frac{P}{\Omega_{\rm K}^2\rho^2}\frac{\partial^2\rho}{\partial z^2}\Bigg|_{z\to 0}=-\frac{1}{\gamma}\left(1+\frac{P}{\Omega_{\rm K}^2\rho}\frac{\partial^2 s}{\partial z^2}\Bigg|_{z\to 0}\right).
\label{eq:term}
\ea     


\bibliographystyle{apj}
\bibliography{references}

\begin{thebibliography}{}
\expandafter\ifx\csname natexlab\endcsname\relax\def\natexlab#1{#1}\fi

\bibitem[{{Bai} \& {Stone}(2013)}]{Bai13}
{Bai}, X.-N., \& {Stone}, J.~M. 2013, \apj, 767, 30

\bibitem[{{Balbus}(2003)}]{Balbus}
{Balbus}, S.~A. 2003, \araa, 41, 555

\bibitem[{{Belyaev} {et~al.}(2012){Belyaev}, {Rafikov}, \& {Stone}}]{Belyaev1}
{Belyaev}, M.~A., {Rafikov}, R.~R., \& {Stone}, J.~M. 2012, \apj, 760, 22

\bibitem[{{Belyaev} {et~al.}(2013{\natexlab{a}}){Belyaev}, {Rafikov}, \&
  {Stone}}]{Belyaev2}
---. 2013{\natexlab{a}}, \apj, 770, 67

\bibitem[{{Belyaev} {et~al.}(2013{\natexlab{b}}){Belyaev}, {Rafikov}, \&
  {Stone}}]{Belyaev3}
---. 2013{\natexlab{b}}, \apj, 770, 68

\bibitem[{{B{\'e}thune} {et~al.}(2016){B{\'e}thune}, {Lesur}, \&
  {Ferreira}}]{Bethune}
{B{\'e}thune}, W., {Lesur}, G., \& {Ferreira}, J. 2016, ArXiv e-prints,
  arXiv:1612.00883

\bibitem[{{Brownlee} {et~al.}(2006){Brownlee}, {Tsou}, {Al{\'e}on},
  {Alexander}, {Araki}, {Bajt}, {Baratta}, {Bastien}, {Bland}, {Bleuet},
  {Borg}, {Bradley}, {Brearley}, {Brenker}, {Brennan}, {Bridges}, {Browning},
  {Brucato}, {Bullock}, {Burchell}, {Busemann}, {Butterworth}, {Chaussidon},
  {Cheuvront}, {Chi}, {Cintala}, {Clark}, {Clemett}, {Cody}, {Colangeli},
  {Cooper}, {Cordier}, {Daghlian}, {Dai}, {D'Hendecourt}, {Djouadi},
  {Dominguez}, {Duxbury}, {Dworkin}, {Ebel}, {Economou}, {Fakra}, {Fairey},
  {Fallon}, {Ferrini}, {Ferroir}, {Fleckenstein}, {Floss}, {Flynn}, {Franchi},
  {Fries}, {Gainsforth}, {Gallien}, {Genge}, {Gilles}, {Gillet}, {Gilmour},
  {Glavin}, {Gounelle}, {Grady}, {Graham}, {Grant}, {Green}, {Grossemy},
  {Grossman}, {Grossman}, {Guan}, {Hagiya}, {Harvey}, {Heck}, {Herzog},
  {Hoppe}, {H{\"o}rz}, {Huth}, {Hutcheon}, {Ignatyev}, {Ishii}, {Ito}, {Jacob},
  {Jacobsen}, {Jacobsen}, {Jones}, {Joswiak}, {Jurewicz}, {Kearsley}, {Keller},
  {Khodja}, {Kilcoyne}, {Kissel}, {Krot}, {Langenhorst}, {Lanzirotti}, {Le},
  {Leshin}, {Leitner}, {Lemelle}, {Leroux}, {Liu}, {Luening}, {Lyon},
  {MacPherson}, {Marcus}, {Marhas}, {Marty}, {Matrajt}, {McKeegan}, {Meibom},
  {Mennella}, {Messenger}, {Messenger}, {Mikouchi}, {Mostefaoui}, {Nakamura},
  {Nakano}, {Newville}, {Nittler}, {Ohnishi}, {Ohsumi}, {Okudaira},
  {Papanastassiou}, {Palma}, {Palumbo}, {Pepin}, {Perkins}, {Perronnet},
  {Pianetta}, {Rao}, {Rietmeijer}, {Robert}, {Rost}, {Rotundi}, {Ryan},
  {Sandford}, {Schwandt}, {See}, {Schlutter}, {Sheffield-Parker},
  {Simionovici}, {Simon}, {Sitnitsky}, {Snead}, {Spencer}, {Stadermann},
  {Steele}, {Stephan}, {Stroud}, {Susini}, {Sutton}, {Suzuki}, {Taheri},
  {Taylor}, {Teslich}, {Tomeoka}, {Tomioka}, {Toppani},
  {Trigo-Rodr{\'{\i}}guez}, {Troadec}, {Tsuchiyama}, {Tuzzolino}, {Tyliszczak},
  {Uesugi}, {Velbel}, {Vellenga}, {Vicenzi}, {Vincze}, {Warren}, {Weber},
  {Weisberg}, {Westphal}, {Wirick}, {Wooden}, {Wopenka}, {Wozniakiewicz},
  {Wright}, {Yabuta}, {Yano}, {Young}, {Zare}, {Zega}, {Ziegler}, {Zimmerman},
  {Zinner}, \& {Zolensky}}]{Brownlee}
{Brownlee}, D., {Tsou}, P., {Al{\'e}on}, J., {et~al.} 2006, Science, 314, 1711

\bibitem[{{Canup}(2004)}]{Canup}
{Canup}, R.~M. 2004, \araa, 42, 441

\bibitem[{{Chiang} \& {Goldreich}(1997)}]{Chiang}
{Chiang}, E.~I., \& {Goldreich}, P. 1997, \apj, 490, 368

\bibitem[{{Ciesla}(2007)}]{Ciesla2007}
{Ciesla}, F.~J. 2007, Science, 318, 613

\bibitem[{{Ciesla}(2009)}]{Ciesla2009}
---. 2009, Icarus, 200, 655

\bibitem[{{Flock} {et~al.}(2011){Flock}, {Dzyurkevich}, {Klahr}, {Turner}, \&
  {Henning}}]{Flock}
{Flock}, M., {Dzyurkevich}, N., {Klahr}, H., {Turner}, N.~J., \& {Henning}, T.
  2011, \apj, 735, 122

\bibitem[{{Fromang} {et~al.}(2011){Fromang}, {Lyra}, \& {Masset}}]{Fromang}
{Fromang}, S., {Lyra}, W., \& {Masset}, F. 2011, \aap, 534, A107

\bibitem[{{Gardiner} \& {Stone}(2008)}]{athena}
{Gardiner}, T.~A., \& {Stone}, J.~M. 2008, Journal of Computational Physics,
  227, 4123

\bibitem[{{Gehrz} {et~al.}(1998){Gehrz}, {Truran}, {Williams}, \&
  {Starrfield}}]{Gehrz}
{Gehrz}, R.~D., {Truran}, J.~W., {Williams}, R.~E., \& {Starrfield}, S. 1998,
  \pasp, 110, 3

\bibitem[{{Hartmann} \& {Davis}(1975)}]{Hartmann}
{Hartmann}, W.~K., \& {Davis}, D.~R. 1975, Icarus, 24, 504

\bibitem[{{Hirose} {et~al.}(2006){Hirose}, {Krolik}, \& {Stone}}]{Hirose06}
{Hirose}, S., {Krolik}, J.~H., \& {Stone}, J.~M. 2006, \apj, 640, 901

\bibitem[{{Hughes} \& {Armitage}(2010)}]{Hughes}
{Hughes}, A.~L.~H., \& {Armitage}, P.~J. 2010, \apj, 719, 1633

\bibitem[{{Illarionov} \& {Sunyaev}(1975)}]{Illarionov}
{Illarionov}, A.~F., \& {Sunyaev}, R.~A. 1975, \aap, 39, 185

\bibitem[{{Jacquet}(2013)}]{Jacquet}
{Jacquet}, E. 2013, \aap, 551, A75

\bibitem[{{Kley} \& {Lin}(1992)}]{Kley1992}
{Kley}, W., \& {Lin}, D.~N.~C. 1992, \apj, 397, 600

\bibitem[{{Kluzniak} \& {Kita}(2000)}]{Kluzniak}
{Kluzniak}, W., \& {Kita}, D. 2000, ArXiv Astrophysics e-prints,
  astro-ph/0006266

\bibitem[{{Lynden-Bell} \& {Pringle}(1974)}]{lynden-bell_1974}
{Lynden-Bell}, D., \& {Pringle}, J.~E. 1974, \mnras, 168, 603

\bibitem[{{Miller} \& {Stone}(2000)}]{Miller2000}
{Miller}, K.~A., \& {Stone}, J.~M. 2000, \apj, 534, 398

\bibitem[{{Narayan} \& {Popham}(1993)}]{Narayan}
{Narayan}, R., \& {Popham}, R. 1993, \nat, 362, 820

\bibitem[{{Pahlevan} \& {Stevenson}(2007)}]{Pahlevan}
{Pahlevan}, K., \& {Stevenson}, D.~J. 2007, Earth and Planetary Science
  Letters, 262, 438

\bibitem[{{Papaloizou} \& {Lin}(1995)}]{papaloizou_1995}
{Papaloizou}, J.~C.~B., \& {Lin}, D.~N.~C. 1995, \araa, 33, 505

\bibitem[{{Popham} \& {Narayan}(1995)}]{Popham2}
{Popham}, R., \& {Narayan}, R. 1995, \apj, 442, 337

\bibitem[{{Popham} {et~al.}(1993){Popham}, {Narayan}, {Hartmann}, \&
  {Kenyon}}]{Popham1}
{Popham}, R., {Narayan}, R., {Hartmann}, L., \& {Kenyon}, S. 1993, \apjl, 415,
  L127

\bibitem[{{Pringle}(1991)}]{Pringle}
{Pringle}, J.~E. 1991, \mnras, 248, 754

\bibitem[{{Rafikov}(2013)}]{Rafikov2013}
{Rafikov}, R.~R. 2013, \apj, 774, 144

\bibitem[{{Rafikov}(2016{\natexlab{a}})}]{Rafikov2016a}
---. 2016{\natexlab{a}}, \apj, 827, 111

\bibitem[{{Rafikov}(2016{\natexlab{b}})}]{Rafikov2016b}
---. 2016{\natexlab{b}}, \apj, 830, 7

\bibitem[{{Regev} \& {Gitelman}(2002)}]{Regev}
{Regev}, O., \& {Gitelman}, L. 2002, \aap, 396, 623

\bibitem[{{Rozyczka} {et~al.}(1994){Rozyczka}, {Bodenheimer}, \&
  {Bell}}]{Rozhichka}
{Rozyczka}, M., {Bodenheimer}, P., \& {Bell}, K.~R. 1994, \apj, 423, 736

\bibitem[{{Shakura} \& {Sunyaev}(1973)}]{shakura_1973}
{Shakura}, N.~I., \& {Sunyaev}, R.~A. 1973, \aap, 24, 337

\bibitem[{{Siemiginowska}(1988)}]{Polka}
{Siemiginowska}, A. 1988, {\rm Acta Astronomica}, 38, 21

\bibitem[{{Stoll} \& {Kley}(2014)}]{Stoll2014}
{Stoll}, M.~H.~R., \& {Kley}, W. 2014, \aap, 572, A77

\bibitem[{{Stoll} \& {Kley}(2016)}]{Stoll2016}
---. 2016, \aap, 594, A57

\bibitem[{{Stoll} {et~al.}(2017){Stoll}, {Picogna}, \& {Kley}}]{Stoll2017}
{Stoll}, M.~H.~R., {Picogna}, G., \& {Kley}, W. 2017, ArXiv e-prints,
  arXiv:1702.00334

\bibitem[{{Suzuki} \& {Inutsuka}(2014)}]{Suzuki}
{Suzuki}, T.~K., \& {Inutsuka}, S.-i. 2014, \apj, 784, 121

\bibitem[{{Syunyaev} \& {Shakura}(1977)}]{SyunShak}
{Syunyaev}, R.~A., \& {Shakura}, N.~I. 1977, Soviet Astronomy Letters, 3, 138

\bibitem[{{Takeuchi} \& {Lin}(2002)}]{Takeuchi}
{Takeuchi}, T., \& {Lin}, D.~N.~C. 2002, \apj, 581, 1344

\bibitem[{{Truran} \& {Livio}(1986)}]{Truran}
{Truran}, J.~W., \& {Livio}, M. 1986, \apj, 308, 721

\bibitem[{{Urpin}(2003)}]{Urpin2003}
{Urpin}, V. 2003, \aap, 404, 397

\bibitem[{{Urpin} \& {Brandenburg}(1998)}]{Urpin1998}
{Urpin}, V., \& {Brandenburg}, A. 1998, \mnras, 294, 399

\bibitem[{{Urpin}(1984)}]{Urpin}
{Urpin}, V.~A. 1984, \azh, 61, 84

\bibitem[{{Wiechert} {et~al.}(2001){Wiechert}, {Halliday}, {Lee}, {Snyder},
  {Taylor}, \& {Rumble}}]{Wiechert}
{Wiechert}, U., {Halliday}, A.~N., {Lee}, D.-C., {et~al.} 2001, Science, 294,
  345

\bibitem[{{Zhu} \& {Stone}(2017)}]{Zhu}
{Zhu}, Z., \& {Stone}, J.~M. 2017, ArXiv e-prints, arXiv:1701.04627

\bibitem[{{Zolensky} {et~al.}(2006){Zolensky}, {Zega}, {Yano}, {Wirick},
  {Westphal}, {Weisberg}, {Weber}, {Warren}, {Velbel}, {Tsuchiyama}, {Tsou},
  {Toppani}, {Tomioka}, {Tomeoka}, {Teslich}, {Taheri}, {Susini}, {Stroud},
  {Stephan}, {Stadermann}, {Snead}, {Simon}, {Simionovici}, {See}, {Robert},
  {Rietmeijer}, {Rao}, {Perronnet}, {Papanastassiou}, {Okudaira}, {Ohsumi},
  {Ohnishi}, {Nakamura-Messenger}, {Nakamura}, {Mostefaoui}, {Mikouchi},
  {Meibom}, {Matrajt}, {Marcus}, {Leroux}, {Lemelle}, {Le}, {Lanzirotti},
  {Langenhorst}, {Krot}, {Keller}, {Kearsley}, {Joswiak}, {Jacob}, {Ishii},
  {Harvey}, {Hagiya}, {Grossman}, {Grossman}, {Graham}, {Gounelle}, {Gillet},
  {Genge}, {Flynn}, {Ferroir}, {Fallon}, {Ebel}, {Dai}, {Cordier}, {Clark},
  {Chi}, {Butterworth}, {Brownlee}, {Bridges}, {Brennan}, {Brearley},
  {Bradley}, {Bleuet}, {Bland}, \& {Bastien}}]{Zolensky}
{Zolensky}, M.~E., {Zega}, T.~J., {Yano}, H., {et~al.} 2006, Science, 314, 1735

\end{thebibliography}

\end{document}